\documentclass[12pt]{article}
\usepackage{epsfig}
\usepackage{float}

\date{}

\newcommand{\ba}{\begin{array}}
\newcommand{\ea}{\end{array}}
\newcommand{\bd}{\begin{displaymath}}
\newcommand{\ed}{\end{displaymath}}
\newcommand{\bi}{\begin{itemize}}
\newcommand{\ei}{\end{itemize}}
\newcommand{\benu}{\begin{enumerate}}
\newcommand{\eenu}{\end{enumerate}}
\newcommand{\be}{\begin{equation}}
\newcommand{\ee}{\end{equation}}
\newcommand{\bea}{\begin{eqnarray}}
\newcommand{\eea}{\end{eqnarray}}

\begin{document}


\begin{center}

{\bf{\large Atmospheric Neutrinos as a Probe of eV$^2 $-scale}}

{\bf{\large Active-Sterile Oscillations}}

\vspace{0.4cm}
Raj Gandhi$\mbox{}^{a)}$,
\footnote{raj@mri.ernet.in}
~and~
Pomita Ghoshal$\mbox{}^{a)}$
\footnote{pomita.ghoshal@gmail.com}

\vspace{0.2cm}

$\mbox{}^{a)}${\em Harish-Chandra Research Institute, Chhatnag Road, Jhunsi, Allahabad - 211019, India.\\
}



\end{center}

\abstract
The down-going atmospheric $\nu_{\mu}$ and ${\bar{\nu_{\mu}}}$ fluxes can be significantly altered due to the presence of eV$^2$-scale active-sterile oscillations.
We study the sensitivity of  a large Liquid Argon detector and  a large magnetized iron detector
(like the proposed ICAL at INO) to these oscillations. Such oscillations are indicated
by results from LSND, and more recently, from MiniBooNE and from reanalyses of reactor experiments following recent recalculations of reactor fluxes. There are other tentative indications of the presence of sterile states in both the $\nu$ and ${\bar{\nu}}$ sectors as well.
Using the allowed sterile parameter ranges in a 3+1 mixing framework in order to test 
these results,
we perform a fit assuming active-sterile oscillations 
in both the muon neutrino and antineutrino sectors,
and compute
oscillation exclusion limits  
 using atmospheric down-going muon neutrino and anti-neutrino events. We find that (for both $\nu_{\mu}$ and ${\bar{\nu_{\mu}}}$) a Liquid Argon detector, an ICAL-like detector or a combined analysis of both detectors with an exposure of 1 Mt yr provides significant sensitivity to regions of parameter space in the range  $0.1 < \Delta m^2 < 5$ eV$^2$ 
for $\sin^2 2\Theta_{\mu\mu}\geq 
0.08$. Thus atmospheric neutrino experiments can provide complementary coverage in these regions, improving sensitivity limits 
in combination with bounds from other experiments on these parameters.
We also analyse the bounds using muon antineutrino events only and compare them with the 
results from MiniBooNE.   

\section{Introduction and Motivation}

Some significant recent experimental results in neutrino physics (MiniBooNE \cite{mini}, reactor ${\bar{\nu_e}}$ flux recalculation \cite{mueller,huber}, gallium data \cite{giunti2}, CMB and big bang nucleosynthesis data \cite{mcgg,raffelt,giu}), along with the older LSND \cite{lsnd} result, have provided  evidence of  ${\bar{\nu_\mu}}\rightarrow{\bar{\nu_e}}$ oscillations at values of $L/E\approx 1$ m/MeV,  where $L$ is the baseline and $E$ the neutrino energy.
 This has motivated the addition of one or more sterile anti-neutrino(s) with eV-scale mass(es) to the standard 3-flavour scenario.
While the experimental evidence remains intriguing, no clearly preferred theoretical model
or framework has emerged so far.  
 

These results have prompted a number of global analyses incorporating recent data as well as results from earlier experiments, which  assume either a framework of three light active neutrinos and one ("3+1"), or two ("3+2") sterile neutrinos \cite{maltoni,kara,akhmedov,kopp,Giunti,giunti3, giunti4, giunti5, giunti6,giunti7,Peres:2000ic,Strumia:2002fw,Grimus:2001mn,Maltoni:2001mt,Palazzo:2011rj,GonzalezGarcia:2001uy,Fogli:2000ir,Karagiorgi:2009nb,GonzalezGarcia:2000hs}.

Efforts to clarify the situation include plans to put in a new near detector or re-using the existing MiniBooNE detector at a near baseline\cite{mills}. There is also a proposal to use the ICARUS detector, a Liquid Argon time projection chamber at the CERN-PS to look for sterile neutrinos{\cite{pietropaolo}}. It has also been suggested that a  large future liquid scintillator detector like NO$\nu$A or LENA be coupled with a compact  decay-at-rest source at a short baseline in order to confirm or refute the signals of sterile neutrinos\cite{sanjib}.

In this situation, it is useful to look at experiments which can provide data at  {\it{multiple}} values of  $L$ and $E$, which would unambiguously signal the presence, or absence, of oscillations. It is also desirable to move
 to a class of experiment with backgrounds and uncertainties which are qualitatively different from those encountered in, for instance, MiniBooNE and LSND, or in reactors. 
 
  Motivated by these considerations,  in this paper we study the role a large atmospheric  detector can play in clarifying these issues. Two examples of such upcoming detectors are the  proposed ICAL at INO \cite{ino} and a large Liquid Argon detector \cite{Rubbia:2004tz,Bueno:2007um,Cline:2006st}.  For simplicity, we consider the down-going muon neutrino and muon anti-neutrino events in the energy range $1-20$ GeV and baseline range $10-100$ km.  This provides a  wide band of $L/E$ with values that are relevant to the issues at hand. The lepton charge identification capability of such detectors, if magnetized, lends an edge by allowing discrimination between the neutrinos and anti-neutrinos, compared to a water Cerenkov detector like SuperKamiokande. It allows independent tests of the presence of sterile neutrinos in the muon and anti-muon data samples, with higher statistical significance for the former due to the larger (by approximately a factor of two) neutrino-nucleon charged current cross section{\footnote{It is not our intention here to assume that there is CPT violation, i.e. that $\nu$ and ${\bar{\nu}}$ oscillate differently. Our results below are based on a combined analysis of $\nu + {\bar{\nu}}$ events. Since the detectors in question can identify lepton charge effectively, we also provide results for ${\bar{\nu}}$ alone, for comparison with MiniBooNE and LSND, which see a signal predominantly in ${\bar{\nu}}$.}}.   
 
In the parameter range under study, strong sterile parameter constraints already exist from CDHS \cite{cdhs}, CCFR \cite{ccfr}, MINOS{\cite{minos1,smirnov,giunti7}}, SciBooNE/MiniBooNE {\cite{mahn}} and SuperKamiokande atmospheric neutrino data \cite{maltoni,sk,maltoni1}. The present limits are summarized in \cite{giunti5}. Recently, even stronger bounds from MINOS$+$ have been suggested in \cite{minos2}, and the possibility of sterile neutrino information from atmospheric neutrino data in IceCube has been discussed in \cite{barger}.
Less stringent constraints also exist for the $\bar{\nu_{\mu}}$ sector from MiniBooNe{\cite{mini1,mini2}.  
 
In Section 2, we motivate our study of sterile-scale oscillations using downgoing atmospheric muon neutrinos, and give the 
specifications of the two futuristic detectors we have analysed for this purpose. Section 3.1 gives our results for the exclusion limits from these detectors in the $\sin^2 2\Theta_{\mu\mu}-\Delta m^2$ plane, 
comparing them with bounds obtained from the experiments listed above.    

 Since both LSND and MiniBooNE have provided evidence of eV$^2$ oscillations in the $\bar{\nu_{\mu}}$ sector,
we separately investigate in Section 3.2 the expected signal from $\bar{\nu_{\mu}}\rightarrow\nu_x$, 
utilizing the charge identification capability of such detectors, for a comparison with the bounds from MiniBooNE. 
Section 4 summarizes our results and conclusions.

\section{Atmospheric down-going muon neutrinos and the sterile-scale oscillation}

Muon and electron neutrinos  and anti-neutrinos produced in the earth's atmosphere provide a naturally occurring source of large  fluxes spanning  an extensive range of energies and baselines. While the upward-going electron and muon neutrinos with baselines of several thousands of Kms pass through the earth, are influenced by earth matter effects and give good sensitivity to standard 3-flavour neutrino oscillation parameters, the downward-going neutrinos have baselines of about 15 - 130 Kms. With an  energy range between 1-20 GeV (above which fluxes become small), these neutrinos lie  in the L/E range in which oscillations arising from the sterile mass-squared difference may be observed. 


Assuming a 3+1 scheme (one non-standard neutrino with a mass squared difference of $\Delta m^2 (=\Delta m_{41}^2) \sim 1$ eV$^2$), the expressions for the relevant survival and oscillation probabilities with 2-flavour sterile-scale oscillations are \cite{giunti3}:

$$
P_{{{\mu}}{{\mu}}} = 1 - \sin^2 2\Theta_{\mu\mu}~\sin^2 [\Delta m^2 L/4E]
$$   

$$
P_{{{e}}{{\mu}}} = \sin^2 2\Theta_{e\mu}~\sin^2 [\Delta m^2 L/4E]
$$    

where $\Theta_{\mu\mu}$ and $\Theta_{e\mu}$ are given by $\sin^2 2\Theta_{\mu\mu} = 4 |U_{\mu 4}|^2 (1 - |U_{\mu 4}|^2)$, $\sin^2 2\Theta_{e\mu} = 4 |U_{e 4}|^2 |U_{\mu 4}|^2$, and $U$ is the 4 $\times$ 4 antineutrino mixing matrix. {{In the energy and baseline range corresponding to downgoing atmospheric neutrinos, the standard 3-flavour oscillations are highly suppressed and one can perform the analysis using only 2-flavour sterile-scale oscillations to a good approximation.}} From \cite{giunti6}, we list the following best-fit values and 3$\sigma$ ranges of these parameters: 
\begin{eqnarray}
\sin^2 2\Theta_{\mu\mu}^{bf} = 0.083,~~~~~0.01 \leq \sin^2 2\Theta_{\mu\mu} \leq 0.25 \\ \nonumber
\sin^2 2\Theta_{e\mu}^{bf} = 0.0023,~~~~~4 \times 10^{-4} \leq \sin^2 2\Theta_{e\mu} \leq 0.01 \\ \nonumber
(\Delta m^2)^{bf} = 0.9~{\rm{eV^2}},~~~~~0.7 \leq \Delta m^2 \leq 7~{\rm{eV^2}}
\label{param1}
\end{eqnarray}

Here the superscript $bf$ denotes the best-fit values.

\vskip.2cm

We perform our statistical analysis using two kinds of proposed detectors:

\begin{itemize}


\item A large Liquid Argon detector, which can detect 
charged particles with very good resolution over the energy range of MeV to multi GeV,  
with magnetization over a 100 kT
volume with a magnetic field of about 1 Tesla \cite{Ereditato:2005yx}.
We  assume the following
energy resolutions over the  ranges relevant to our
calculations 
\cite{Bueno:2007um}: 
\bea
\sigma_{E_e} = 0.01,~~~~~\sigma_{E_{\mu}} = 0.01 \nonumber \\   
\sigma_{E_{had}} = \sqrt{(0.15)^2/E_{had} + (0.03)^2} \nonumber \\
\sigma_{\theta_e} = 0.03~{\rm{radians}} = 1.72^o,~~~~~\sigma_{\theta_{\mu}} = 0.04~{\rm{radians}} = 2.29^o \nonumber \\   
\sigma_{\theta_{had}} = 0.04~{\rm{radians}} = 2.29^o  
\eea 
where $E_{e}$, $E_{\mu}$ and $E_{had}$ are the lepton and hadron energies in GeV, $\sigma_E$ are the energy resolutions and $\sigma_{\theta}$ are the angular resolutions 
of electrons, muons and hadrons as indicated. 
The energy resolution in terms of the neutrino energy is related to the leptonic and hadronic
energy resolutions as follows:
\be
\sigma_{\nu}/E_{\nu} = {\sqrt{(1-y)^2 (\sigma_{lep} / E_{lep})^2 + y^2 (\sigma_{had}/E_{had})^2}}
\ee
Here the rapidity or the Bjorken scaling variable $y$ is defined as $y = E_{had}/E_{\nu}$, where $E_{\nu} = E_{lep} + E_{had}$ is the energy of the neutrino.
{{The relation $E_{\nu}=E_{lep}+E_{had}$ is exact for quasi-elastic scattering and the closest analytic approximation for DIS scattering. Lepton-neutrino collinearity is assumed in this procedure, and is expected to hold true to a good approximation in the multi-GeV neutrino energy range.}}
Hence the energy resolution in terms of the neutrino energy is given by
\be
\sigma_{E_{\nu}} = {\sqrt{(0.01)^2 + (0.15)^2/y E_{\nu} + (0.03)^2}}
\ee
for both electron and muon neutrinos.
In our computation, we take the average rapidity in the GeV energy region 
to be 0.45 for neutrinos and 0.3 for antineutrinos \cite{Gandhi:1995tf}. 
{{These average rapidity values have been verified to
be accurate using realistic hadron event simulations.}}
The angular resolution of the detector for neutrinos can be worked out to be $\sigma_{\theta_{\nu e}} = 2.8^o$,
$\sigma_{\theta_{\nu\mu}} = 3.2^o$. 
The energy threshold and ranges in which charge identification is feasible  are 
$E_{threshold}  = 800$ MeV for muons and
$E_{electron}  = 1-5$ GeV  for electrons.

\item An iron calorimeter detector like ICAL, which, like the above detector, offers the advantage of muon charge discrimination using magnetization with a field of 1.3 Tesla, allowing a separate observation of atmospheric muon neutrino and anti-neutrino events.  
For this detector, standard resolutions of 10$^o$ in angle and 15$\%$ in energy are assumed. 

\end{itemize}

The muon event rates are a function of both $P_{{{\mu}}{{\mu}}}$ and  $P_{{{e}}{{\mu}}}$, but  $P_{{{e}}{{\mu}}}$ is highly suppressed due to the smallness of the parameter $\sin^2 2\Theta_{e\mu}$. Thus  the downgoing muon event spectrum reflects the behaviour of $P_{{{\mu}}{{\mu}}}$ and shows signatures of the sterile parameters $\Theta_{\mu\mu}$ and $\Delta m^2$. 
In Figure 1, the downgoing muon neutrino distribution with and without two-flavour sterile oscillations is shown as a function of the neutrino energies, integrated over $\cos \theta_z$ bins. 

We assume a 1 Mt yr exposure for both types of detectors, standard detector resolutions as above, and flux uncertainties and systematic errors
incorporated by the pull method \cite{Gonzalez}. The values of uncertainties are chosen as in \cite{Gandhi}:  
flux normalization error 20$\%$, 
flux tilt factor, zenith angle dependence uncertainty 5$\%$, overall cross-section uncertainty 10$\%$, overall systematic uncertainty 5$\%$. 
{{We take a double binning in neutrino energy and zenith angle in the energy range 1-20 GeV and $\cos \theta_z$ range 0.1 to 1.0.   
To be consistent with the detector resolution, 
the bin widths are required to be $\ge$ the resolution widths.
For the above neutrino energy and zenith angle ranges,
this allows us to take 20 energy bins and 18 angle bins for these values
of resolution width. 
The atmospheric fluxes are taken from the 3-dimensional calculation in \cite{Honda:2004yz}.
{{We have taken into account the neutrino production height distribution in the atmosphere \cite{Gaisser:1997eu}.}}

\begin{figure}[t]
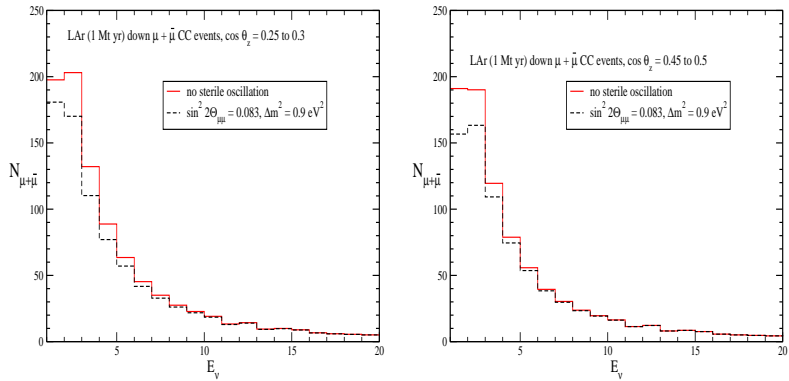

{\centerline
{
\hspace*{1em} \epsfxsize=5cm\epsfysize=5cm
                     \epsfbox{EventdownhistvsEm_LiqAr1Mtyr_costhm0.25to0.3_sterandnoosc_sm10th15E_corr_test.eps}
         \hspace*{0.5ex}
\epsfxsize=5cm\epsfysize=5cm
                      \epsfbox{EventdownhistvsEm_LiqAr1Mtyr_costhm0.45to0.5_sterandnoosc_sm10th15E_corr_test.eps}}}
\hskip 5cm
\caption[]{\footnotesize{LAr downgoing $\mu + {\bar{\mu}}$ event spectrum vs neutrino energy integrated over the 2 specific $\cos \theta_z$ bins with and without 2-flavour sterile-scale oscillations, using best-fit sterile parameter values.}}
\label{fig1a} 
\end{figure}

\section{Exclusion limits with sterile oscillations in the $\nu_{\mu}$ and ${\bar{\nu}_{\mu}}$ event spectra}
One can extract the statistical sensitivity with which experimental-set ups like the ones described above may be able to constrain sterile parameters using the downgoing muon and anti-muon event spectra as the signal.
We perform this study in two stages: 

a) The best exclusion limits possible from this analysis are determined using simultaneously the downgoing $\nu_{\mu}$ and ${\bar{\nu}_{\mu}}$ event spectra for both kinds of detectors and doing a combined fit.

b) In order to test the MiniBooNE/LSND antineutrino results, the atmospheric downgoing ${\bar{\nu}_{\mu}}$ event spectra with sterile oscillations for both kinds of detectors are analysed to determine the sterile parameter bounds, and compared with the bounds from MiniBooNE.

\subsection{Limits with a combined $\nu_{\mu}$, ${\bar{\nu}_{\mu}}$ analysis}


\begin{figure}[t]
{\centerline
{
\hspace*{1em} \epsfxsize=6cm\epsfysize=6cm
                     \epsfbox{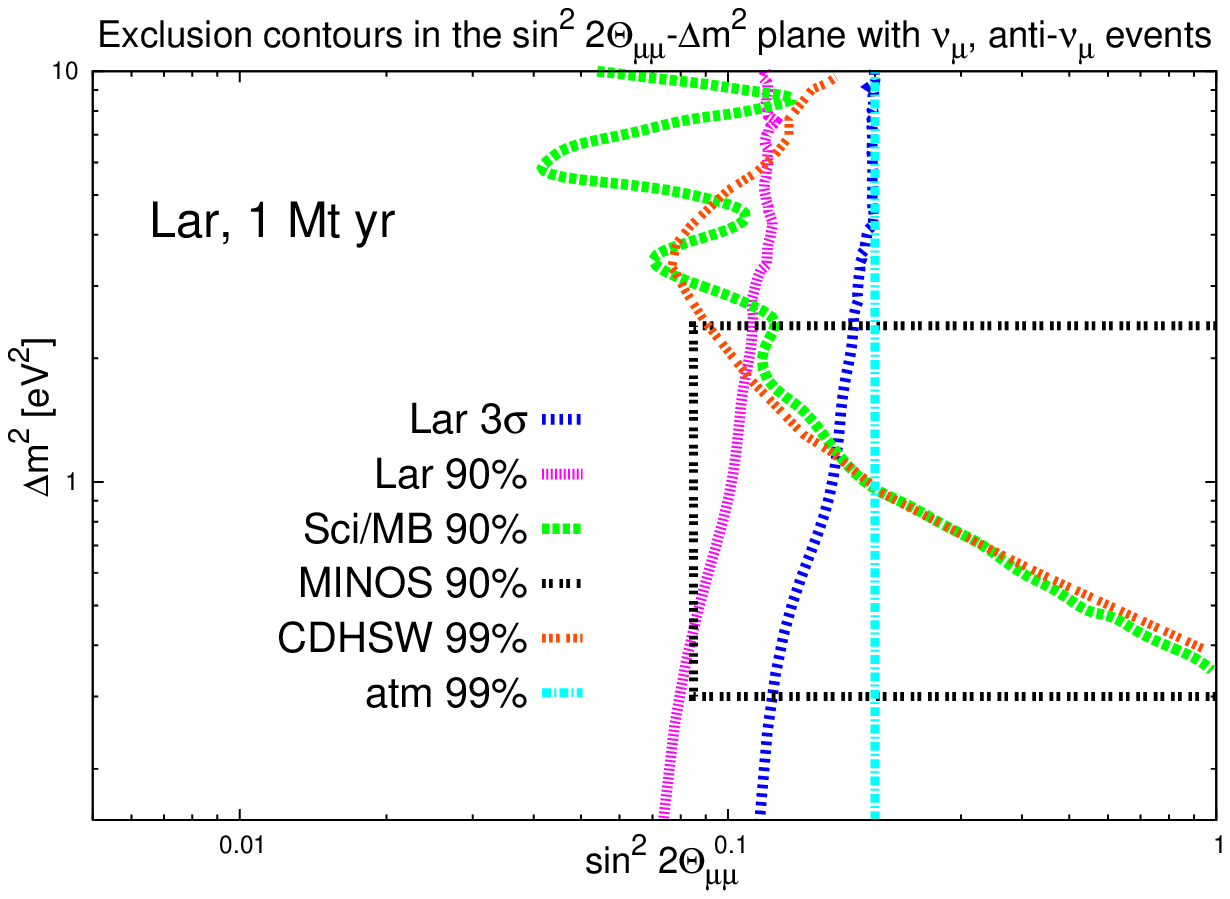}
         \hspace*{0.5ex}
\epsfxsize=6cm\epsfysize=6cm
                      \epsfbox{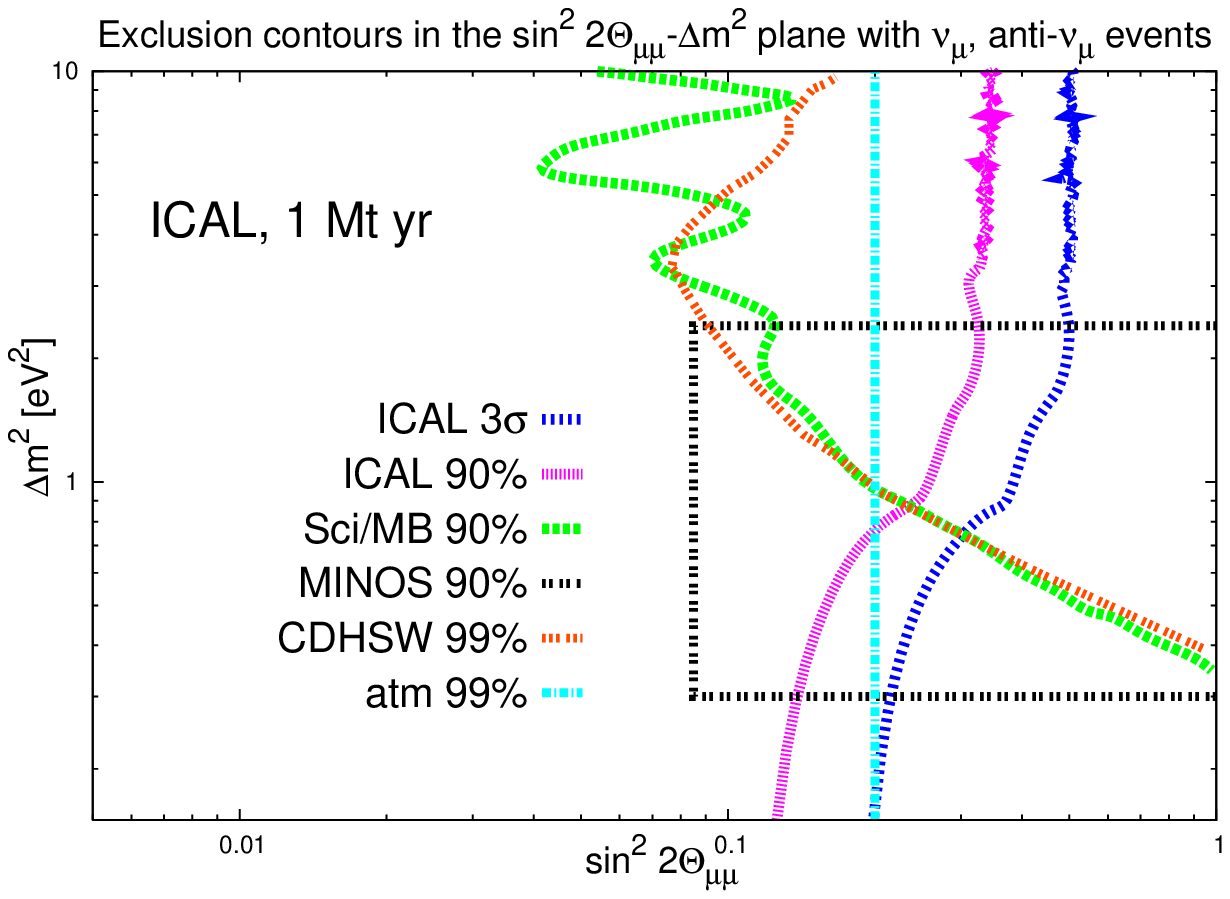}}}
\hskip 5cm
{\bf (a)}
\hskip 7cm
{\bf (b)}
\caption[]{\footnotesize{Exclusion curves  
using (a) Liquid Argon and (b) ICAL downgoing muon and anti-muon events 
with sterile oscillations - Comparison with 99$\%$ c.l. limit from atmospheric 
analysis \cite{giunti5,maltoni}, 90$\%$ limits from SciBooNE/MiniBooNE {\cite{mahn}} 
and MINOS {\cite{minos1}}, 
99$\%$ c.l. limit from CDHSW {\cite{giunti5,cdhs}}.}}
\label{fig1} 
\end{figure}

For this study, 
the sterile oscillation exclusion limits are computed by combining both the atmospheric downgoing muon and antimuon event spectra
with sterile-scale oscillations. 
This involves  taking i) the 'expected' spectrum $N_{th}$, 
in which sterile oscillations are included,
and ii) the 'observed' spectrum $N_{ex}{\rm{(no-osc)}}$, where 'no-osc' indicates no sterile-scale oscillations.
{{Since we expect no differences in the oscillations of neutrinos compared to antineutrinos
in the 3$+$1 model when matter effects are absent, we have assumed equal probabilities for them
for downgoing events.}}
The exclusion limits are presented in Figure 2 for a Liquid Argon detector and an ICAL detector, with an exposure of 1 Mt yr for both. Figure 3 shows the results from a combined analysis of ICAL + Liquid Argon.
The bounds obtained from our analysis are compared with the 99$\%$ c.l. exclusion region from atmospheric neutrino data \cite{giunti5,maltoni}, the 90$\%$ limits from SciBooNE/MiniBooNE \cite{mahn} and MINOS \cite{minos1} and the 99$\%$ limit from the CDHSW disappearance analysis \cite{giunti5,cdhs}.

 With a Liquid Argon detector and an exposure of 1 Mt yr,
regions greater than $\sin^2 2\Theta_{\mu\mu}\sim 0.09$ can be excluded at 90$\%$ c.l. with a combination of muon and anti-muon events over most of the allowed $\Delta m^2$ range.
An ICAL-like detector with a similar exposure gives a weaker 90$\%$ c.l. exclusion bound 
at $\sin^2 2\Theta_{\mu\mu} \sim 0.15$.  
A combined analysis of the two experiments gives a 3$\sigma$ exclusion bound for $\sin^2 2\Theta_{\mu\mu}\geq 0.11$, 
and a 90$\%$ c.l. limit for $\sin^2 2\Theta_{\mu\mu}\geq 0.08$,
which is seen to be an improvement over the earlier bounds obtained from atmospheric neutrinos, as well as those from CDHSW, SciBooNE/MiniBooNE and MINOS, over significant regions of the parameter space. 
{{Note that the sensitivity using this set-up is better in the low $\Delta m^2$ region ($\Delta m^2 < 1$ eV$^2$), part of which lies outside the present global fit range, but our purpose here is to demonstrate that such an experiment is capable of providing significant bounds over the entire parameter space which can contribute to constraining sterile parameters in combination with other experiments in a future global analysis.}} 

\begin{figure}[t]
{\centerline
{
\hspace*{1em} \epsfxsize=6cm\epsfysize=6cm
                     \epsfbox{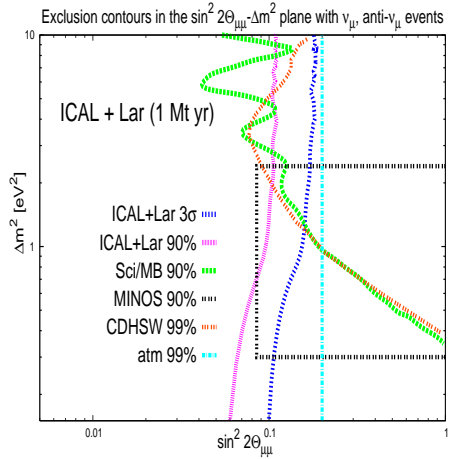}}}
\caption[]{\footnotesize {Same as Figure 2 using ICAL $+$ Liquid Argon downgoing muon and antimuon events. }}
\label{fig2} 
\end{figure}

\begin{figure}[t]
{\centerline
{
\hspace*{1em} \epsfxsize=6cm\epsfysize=6cm
                     \epsfbox{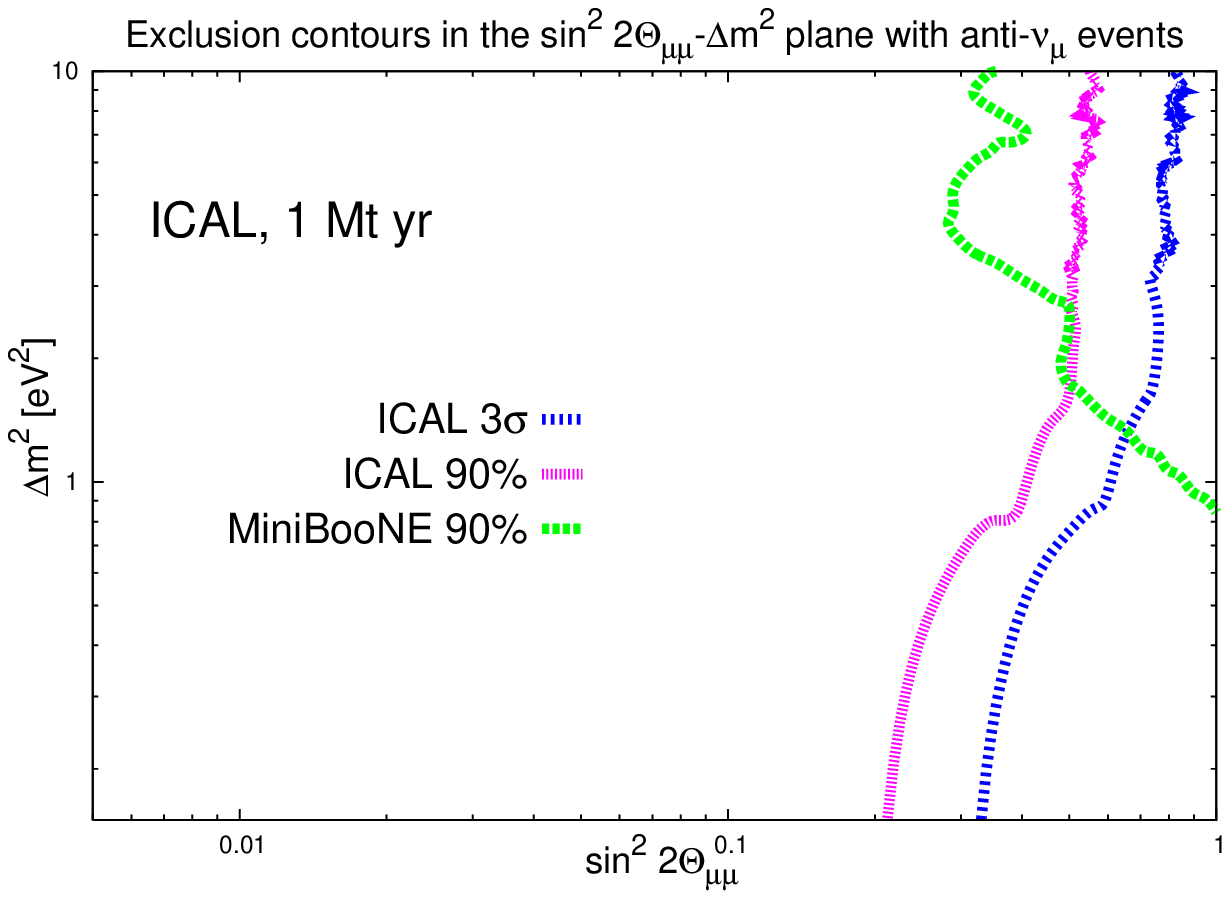}
         \hspace*{0.5ex}
\epsfxsize=6cm\epsfysize=6cm
                      \epsfbox{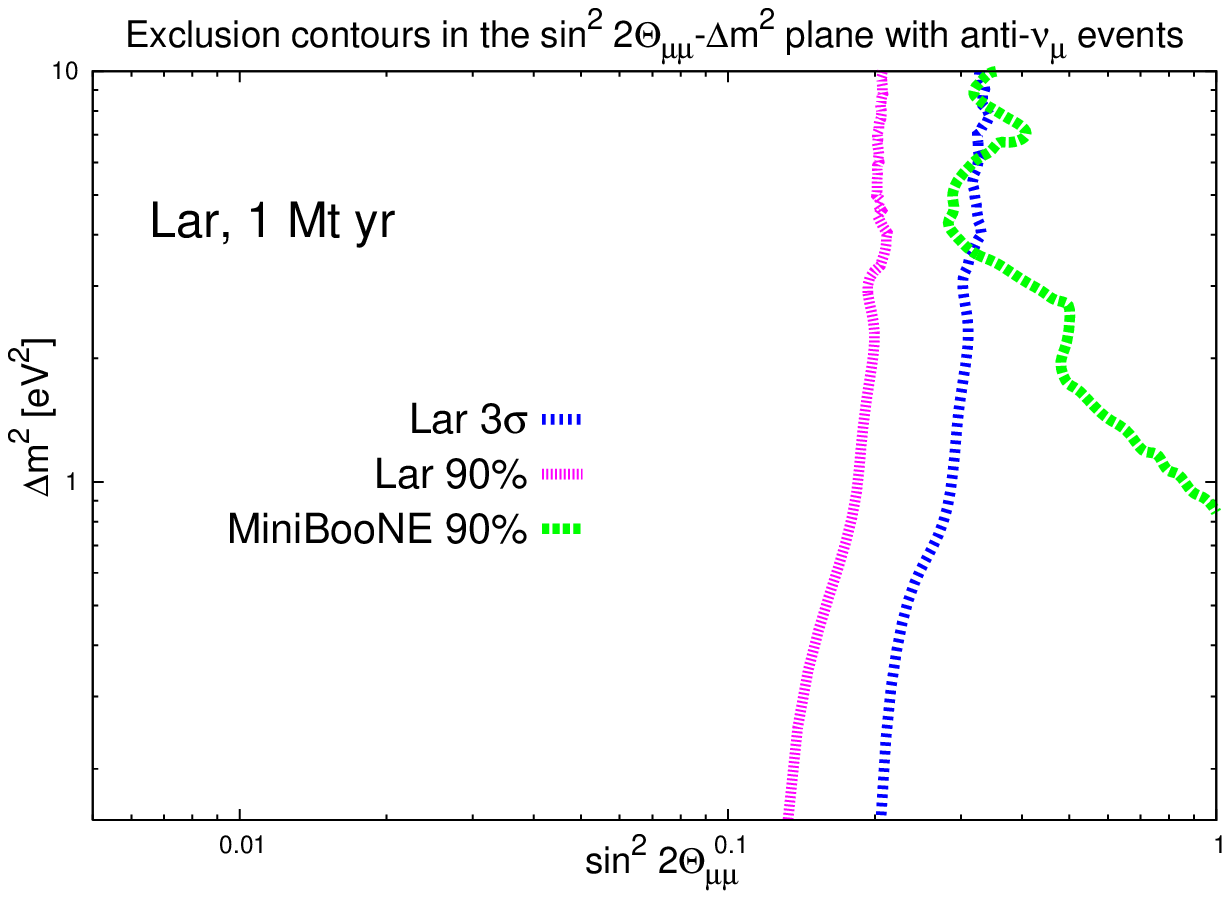}}}
\hskip 5cm
{\bf (a)}
\hskip 7cm
{\bf (b)}
\caption[]{\footnotesize {Exclusion curves using (a) ICAL  and (b) Liquid Argon downgoing anti-muon events with sterile oscillations - Comparison with
90$\%$ exclusion limit from MiniBooNE \cite{mini2} antineutrino analysis.}}
\label{fig3} 
\end{figure}

\subsection{Testing LSND and Miniboone results with sterile oscillations in the ${\bar{\nu}}$ sector}
Here we compute the   
 sterile oscillation exclusion limits using the downgoing antimuon event spectrum for comparison with the results from MiniBooNE/LSND antineutrino data. 
The bounds obtained from this analysis are presented in Figure 4.
The left and right panels correspond to the exclusion bounds for the parameters ${\bar{\Delta m^2}}$ and 
$\sin^2 2{\overline{\Theta}_{\mu\mu}}$ with the downgoing ${\bar{\nu}}_{\mu}$ spectrum for
an ICAL detector and a Liquid Argon detector respectively, both with an exposure of 1 Mt yr.   
Here ${\bar{\Delta m^2}}$ and ${\overline{\Theta}_{\mu\mu}}$ denote the mixing parameters 
for antineutrinos.  
The 90$\%$ exclusion limit from MiniBooNE \cite{mini2} is superimposed on both figures.
It can be seen that this set-up provides a 90$\%$ c.l. exclusion capacity with 
an ICAL-like detector with an exposure of 1 Mt yr for about $\sin^2 2{\overline{\Theta}_{\mu\mu}} > 0.4$ for a range $0.1 < {\bar{\Delta m^2}} < 5$ eV$^2$, and for a Liquid Argon detector with an exposure of 1 Mt yr for about $\sin^2 2{\overline{\Theta}_{\mu\mu}} > 0.15$ for a range $0.1 < {\bar{\Delta m^2}} < 5$ eV$^2$. {{Thus a Liquid Argon detector gives stronger bounds than those from the MiniBooNE antineutrino analysis. }}

\section{Summary}

In this paper, we have studied the  possible sensitivity of atmospheric neutrino data in a large magnetized iron calorimeter detector
like the proposed ICAL at INO and a large Liquid Argon detector, to eV$^2$-scale active-sterile oscillations. Such detectors are capable of distinguishing lepton charge and hence discriminating between neutrino and antineutrino events. 
With the present sterile parameter ranges 
in a 3+1 mixing framework, 
down-going atmospheric ${{\nu}_\mu}$ and ${\bar{\nu}_\mu}$ events can show signatures of eV$^2$-scale oscillations, due to their suitable energy and baseline range (neutrinos with multi-GeV energies and baselines ranging from about 10 to 100 Kms). Our analysis has been done in two parts:

\begin{itemize}

\item To be consistent with homogeneity in $\nu-{\bar{\nu}}$ behaviour in the 3+1 scenario,
we assumed identical active-sterile oscillations in both the muon neutrino and antineutrino sectors
and derived active-sterile oscillation exclusion limits (Figures 2 and 3), comparing them with
the limits obtained from \cite{mahn,minos1,giunti5,giunti6,cdhs,maltoni}.


a) With a Liquid Argon detector (1 Mt yr),
regions greater than $\sin^2 2\Theta_{\mu\mu}\sim 0.09$ can be excluded at 90$\%$ c.l. with a combination of muon and anti-muon events, over most of the $\Delta m^2$ range. 
A 3$\sigma$ exclusion bound is possible for $\sin^2 2\Theta_{\mu\mu}\sim 0.13$.

b) With an ICAL detector (1 Mt yr),
a weaker 90 $\%$ c.l. exclusion bound is obtained at $\sin^2 2\Theta_{\mu\mu} \sim  
0.15$ with a combination of muon and anti-muon events.

c) With a combined analysis of ICAL (1 Mt yr) and Liquid Argon (1 Mt yr), a 90$\%$ c.l. exclusion limit is obtained for 
$\sin^2 2\Theta_{\mu\mu} \geq 0.08$ and a 3$\sigma$ bound for $\sin^2 2\Theta_{\mu\mu} \geq  
0.11$, which compares favorably with present limits from CDHSW, MINOS, MiniBooNE and atmospheric neutrinos.



\item For testing the predictions of LSND and MiniBooNE, 
we performed a fit with active-sterile oscillations in the muon antineutrino sector, comparing the results with
the exclusion limits from MiniBooNE (Figure 4),
and derived the following 90$\%$ c.l. exclusion bounds: 

a) an ICAL-like detector with an exposure of 1 Mt yr for about $\sin^2 2{\overline{\Theta}_{\mu\mu}} > 0.4$ for a range $0.1 < {\bar{\Delta m^2}} < 5$ eV$^2$, 

b) A Liquid Argon detector with an exposure of 1 Mt yr for about $\sin^2 2{\overline{\Theta}_{\mu\mu}} > 0.15$ for a range $0.1 < {\bar{\Delta m^2}} < 5$ eV$^2$.  

\end{itemize}

The limits for both detectors from an exposure of 1 Mt yr may be accessible 
in a time-frame of about 10-15 years. 

In conclusion, a down-going event analysis using large future atmospheric  detectors may be helpful in providing 
significant complementary constraints on the sterile parameters, which can strengthen existing bounds 
when combined with other experimental signatures of sterile-scale oscillations.  
{{Such a set-up exhibits better sterile oscillation sensitivity in the low $\Delta m^2$ region ($\Delta m^2 < 1$ eV$^2$).}}
Evidence (or the lack of it) from such detectors has the advantage of originating in a sector which is different from those currently providing clues pointing to the existence of sterile neutrinos (i.e short-baseline experiments). Additionally, it permits access to a wide-band of $L/E$, which is important if oscillatory behaviour is to be unambiguously tested.
   
 \section{Acknowledgements} 
 We acknowledge long-standing collaborations on INO/ICAL physics with Srubabati Goswami and S. Uma Sankar, which have informed this work. RG would like to acknowledge the hospitality of the 
 Cern Theory Division, where this work was initiated. He is also grateful for financial support from the DAE XI Plan Neutrino project. PG thanks S.T. Petcov for useful discussions.
This work was supported in part by the
INFN program on "Astroparticle Physics" (PG).

\end{document}